# Generative Software Development

Bernhard Rumpe, Martin Schindler, Steven Völkel, Ingo Weisemöller
Software Engineering
RWTH Aachen University, Germany
http://www.se-rwth.de/

## ABSTRACT

Generation of software from modeling languages such as UML and domain specific languages (DSLs) has become an important paradigm in software engineering. In this contribution, we present some positions on software development in a model based, generative manner based on home grown DSLs as well as the UML. This includes development of DSLs as well as development of models in these languages in order to generate executable code, test cases or models in different languages.

Development of formal DSLs contains concepts of metamodels or grammars (syntax), context conditions (static analysis and quality assurance) as well as possibilities to define the semantics of a language. The growing number and complexity of DSLs is addressed by concepts for the modular and compositional development of languages and their tools. Moreover, we introduce approaches to code generation and model transformation. Finally, we give an overview of the relevance of DSLs for various steps of software development processes.

## 1. INTRODUCTION

Languages are a key concern for software development. This does not only apply to programming languages for the implementation, but also to natural languages for requirements specification and documentation, and to modeling languages such as UML. In addition to these languages, domain specific languages (DSLs) have recently become increasingly important to various activities in software development processes. In order to develop a formal language and to integrate it into a software development process, the following steps need to be performed [1, 7].

First of all, the language has to be defined precisely. This includes a description of the valid words of the language, which is determined by its syntax and by context conditions. These are often described by means of context free grammars, attribute grammars, symbol tables and constraints. The language definition also includes a description of the semantics of the language [4]. This is often implemented by means of model transformations or code generation.

The language can then be used in other software development processes. Thus, developers need tools to describe and transform models in the language, and the process has to be adopted to the usage of the new language. Moreover, measures for quality assurance of documents in the language are required.

In the remainder of this contribution, we give a brief overview of preliminary considerations about the use of a DSL, important methods and techniques that are crucial for the definition of the syntax and context conditions, and the implementation of generators and transformations. We also consider modularization concepts for DSL definitions, quality assurance and the integration of a domain specific language into software development processes.

## 2. PRELIMINARIES FOR DSL USAGE

The development of a domain specific language and the corresponding tools is a software development process itself, which may be expensive and error-prone. Therefore, the introduction of a DSL is particularly useful in the development of large and complex products [3]. In smaller development processes, the improvements in terms of efficiency and software quality may not be sufficient to compensate the initial costs that are caused by the DSL development.

## 3. SYNTAX AND CONTEXT CONDITIONS

The syntax of a language can be defined a context free grammar. This includes the concrete syntax, i.e. the external representation of the language to the user, and the abstract syntax, i.e. the internal representation of the corresponding data structures. There are several compiler generating tools available, which can generate the implementation of the abstract syntax, parsers, pretty printers and editors for languages from their grammars [1, 2]. However, some restrictions for the set of valid words in a language cannot be defined in such a grammar, because they are context sensitive. Typical examples are validity of identifiers, type safety or acyclicity of inheritance, package or composition hierarchies. For these restrictions, additional analyses are required, which are usually implemented in a constraint language (inside the grammar) or as an external program linked against the classes of the abstract syntax.

As an alternative to grammars, metamodels can be used to define the abstract syntax of a language. As with grammars, we often encounter a combination of metamodels and constraint languages. In order to define the concrete syntax

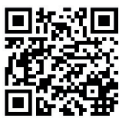

of a language, metamodel based tools often rely on editor frameworks.

## 4. GENERATORS

Generators are tools that transform models to software. They can be used to implement the semantic mapping of a language. We can distinguish betwenn model-to-text transformations, which transform a model to a document in a textual language, and model-to-model transformations, which transform it into a model in a different modeling language. For both kinds of transformations, it is crucial that the target language has precisely defined semantics. In case of executable models, the target language is often a general purpose language (GPL) such as Java or C++, and the runtime semantics of the source model are the runtime semantics of the generated code.

Model-to-Model transformations are formally founded on graph grammars and graph transformations. They are usually implemented in dedicated model transformation languages (which can in turn be considered to be domain specific languages).

Most model-to-text-transformations are implemented by means of template languages such as Velocity or Freemarker. Model transformations can be executed locally on the machine of a developer, or remote as a transformation service. Software engineering services for transformations reduce the technical efforts that are necessary to use a DSL.

## 5. MODULAR DSL DEFINITIONS

The growing number and complexity of DSLs results in the necessity to develop new languages efficiently [8]. One obvious step towards this goal is to reuse existing artifacts during language development. Inheritance concepts similar to those known from object oriented programming can also be applied to grammars or production rules inside grammars. Moreover, it is possible to embed one language into another, such as known for languages like Java Server Pages, which embed Java Code into HTML pages [5].

## 6. QUALITY ASSURANCE

As in all software development processes, quality assurance is important when using domain specific languages. In contrast to software development with previously existing languages, quality assurance is not limited to the documents in the development process of system under development, but is also required for the development of the DSL itself.

Quality aspects of a language include the language definition as well as the tool infrastructure provided to the users of the language. Quality of language definitions is difficult to ensure, although guidelines and reviews are measures that can be transferred from programming to language development. The appropriate measures for quality assurance of the tool infrastructure depend massively on the development method for the tools. If most parts of the tools are generated, the quality of the language definition is important. If the tools are handcoded, concepts for quality assurance known from traditional software development are appropriate.

The introduction of a domain specific language is accompanied by the question how to assure the quality of instances of this language, i.e. the models. Not all models that are valid w.r.t. to the syntax and context conditions of the language are of appropriate quality. Naming conventions may help to make the documents easier to read and more comprehensible. The same goes for indentation guidelines, as well as for conventions on comments, which may prescribe what information should be contained in the comments of a document and which (natural) language should be used.

Although modeling languages need to be quality assured as described above, they may on the other hand contribute to quality assurance for a system under development [6]. If tests are used for quality assurance, input and output data of test cases can be described as models. In this case, a test case consists of an input model, the implementation of the runtime behavior of the test, which may be written in a GPL, and an output model that describes the expected result of the test. The test case fails if and only if the actual output differs from the output model. Modeled test cases are very appropriate for test driven development, because they allow for a rapid specification of high-quality tests, which will result in faster development and excellent test coverage.

## 7. DSLS IN SOFTWARE ENGINEERING

Modeling languages and domain specific languages have recently found their way into most activities of the software lifecycle. In the requirements and analysis phase, algebraic specification languages have been used for several years. Recently, requirements specification languages that are close to natural languages have been introduced in this area. Architectural description languages and UML play an important role in system design. Matlab/Simulink is a wide spread language for the implemenetation of electronic control units in automotive industry. The model-based specification of test cases has been described in the previous chapter. Therefore, the development of high-quality languages and language instances can contribute significantly to more efficient and valuable software systems.